# Mechanically Stimulated Domain Re-arrangement in Cryogenic Magnetic Shielding Materials

Anthony C. Crawford, *Fermilab*

*Abstract*—Room temperature properties and behavior of several types of cryogenic magnetic shielding materials are measured and reported here. Large changes in the effective relative permeability are observed when the materials are perturbed with a relatively small mechanical stimulation. The change in permeability is a reversible effect.

## I. Introduction

THE importance of cryogenic magnetic shielding material has increased significantly since the introduction of "High-Q" nitrogen doped niobium as a practical surface for superconducting RF (SRF) cavities intended for particle accelerator and storage ring use. The RF surface resistance of "N-doped" cavities is more sensitive to trapped magnetic flux than the surface of pure niobium cavities [1], and requires extremely effective ambient field attenuation from magnetic shielding.

This report summarizes measurements made on sample cylinders of shielding material at room temperature with the aim of characterizing their ability to attenuate low (~50 µTesla) steady state magnetic field.

## II. Shielding material

Materials from three suppliers take part in this study. They are listed in Table 1. All are heat treated to increase their permeability at cryogenic temperatures.

TABLE I
SHIELDING MATERIALS

| SUPPLIER | MATERIAL |
|---|---|
| Ad-Vance Magnetics | CP-EXP-1115 |
|  | CP-EXP-1184 |
| Amuneal Mfg. Corp | A4K |
| Sekels, GMBH | Cryoperm10 |

Cryogenic shielding material in a cavity cryomodule can be at temperatures usually in the range 1.8K≤T≤80K. T908he permeability of commercially available cryogenic magnetic shielding material has been found to decrease by as much as a factor of two when cooled from 300K to 4K.

### A. Material Composition

The chemical composition of the materials was analyzed using a Bruker S1 Turbo$^{SD}$ spectrometer. Results are shown in Table II. Components that measured less than 0.4 % are not included in the table.

TABLE II
SHIELDING MATERIAL % COMPOSITION

| ELEMENT | 1115 | 1184 | A4K | CRYOPERM10 |
|---|---|---|---|---|
| Cr | - | 2.3 | - | - |
| Mn | 0.4 | 0.5 | 0.5 | 0.5 |
| Fe | 15.3 | 16.8 | 15.0 | 15.8 |
| Ni | 78.6 | 75.2 | 79.7 | 76.3 |
| Cu | - | 5.1 | - | 5.0 |
| Mo | 5.1 | - | 4.4 | 2.4 |

It is notable that both the 1115 material and A4K are chemically similar to Permalloy80, invented at Bell Telephone Laboratories in 1914. The antiquity of Permalloy80 is such that Alexander Graham Bell would still live for six years after its introduction!

### B. Material Heat Treatment

All samples were heat treated by Ad-Vance after being formed and welded into the shape of tubes. A generic heat treatment for cryogenic magnetic shields is described in reference [2]: "2-4 h at 1,100C in vacuum; 2 h cooling to 500C and then with helium or argon gas to room temperature; annealing under vacuum for 0.5 h at 570C; slow cooling to 470C and annealing there for 2 h; fast He or AR gas cooling to room temperature."

## III. Attenuation Measurements

The method used for this report is to measure the attenuation of applied magnetic field in the axial direction of open cylinders. Obtaining adequate axial attenuation, or as it is sometimes called, longitudinal attenuation, is the most difficult task facing the SRF cryomodule magnetic shield designer. This is because the attenuation of a cylinder along its axis decreases as the square of the ratio of diameter to length, and most SRF cavity shields are long, narrow cylinders. The intention of the present study is to be practical, with a goal of making measurements that lead to useful design parameters for magnetic shields, rather than those that are more relevant for investigation of the underlying physics.

*A. Test Geometry*

The axis of the sample cylinder was oriented in the vertical direction. All magnetic field in the direction transverse to the cylinder axis was cancelled with a large set of external Helmholtz coils. Cancellation was done when the sample cylinder and all magnetic material (except for the magnetic sensor) was not present within the active volume of the Helmholtz coils. The field in the vertical direction was adjusted with external coils to be 50.0 µTesla when there was no magnetic sample material present. A Bartington single axis fluxgate magnetometer was located at the geometric center of the tube. The orientation of the sensor axis was in the vertical direction. The Bartington sensor was a model suited for cryogenic use. The accuracy of the fluxgate with its readout electronics is ±1 percent of the measured field, dominated by uncertainty in the scaling coefficient.

*B. Measured Attenuation*

Attenuated field values ($B_{axial}$) are listed in Table III, where length and diameter measurements are listed in millimeters and attenuated field is in µTesla. The axial attenuation factor is calculated by dividing the ambient axial field (50.0 µTesla) by the attenuated field value at the geometric center of each sample. All measurements were taken at room temperature (~20C). Remanent field in the samples was adjusted to zero before measurements were taken. The thickness of all tested materials was 1 millimeter.

TABLE III
ATTENUATED FIELD MEASUREMENTS

| SAMPLE | INNER DIAMETER | LENGTH | $B_{AXIAL}$ (µTESLA) | ATTENUATION FACTOR |
|---|---|---|---|---|
| 1115 | 25.1 | 127.0 | 0.081 | 617 |
| 1184 | 22.8 | 127.0 | 0.168 | 298 |
| A4K | 25.3 | 127.0 | 0.090 | 556 |
| Cryoperm10 | 25.4 | 127.0 | 0.178 | 281 |

## IV. ESTIMATED RELATIVE PERMEABILITY

A two dimensional finite element magnetic modeling program was used to obtain a value for relative permeability that best predicts the measured value for attenuation in Table III [3]. Permeability is forced to have a constant value over the range of magnetic field inside the shielding material. The range of field inside the metal of the shield is approximately 1000 µTesla to 8000 µTesla. Field values this small justify the use of a linear B-H curve segment for the model. Table IV lists the calculated relative permeability. The model featured the specific geometry of each test cylinder immersed in an applied axial field of 50.0 µTesla.

TABLE IV
CALCULATED CONSTANT RELATIVE PERMEABILITY

| SAMPLE | RELATIVE PERMEABILITY |
|---|---|
| 1115 | 98,500 |
| 1184 | 47,500 |
| A4K | 88,500 |
| Cryoperm10 | 44,000 |

## V. MEASUREMENTS WITH MECHANICAL STIMULATION

In the course of taking measurements, it was observed that small mechanical perturbations in the form of applied mechanical shock would change the value of attenuated field. The trend was that larger perturbations resulted in larger changes in the field and that attenuation always increased with the application of shock. In order to quantify this effect, each sample was tapped with a plastic hammer weighing 3.3 grams while the sample was immersed in a 50.0 µTesla axial applied field. The estimated maximum kinetic energy of the hammer was $1 \times 10^{-4}$ joule. The mass of each sample was approximately 90 grams. Since less than the full kinetic energy of the hammer was transferred to the sample, the added energy to the sample was less than 1 millijoule per kilogram per hammer strike. It should be noted that the hammer strikes were executed by hand, and not by a machine with precise control and repeatability. The impact energy per strike was approximately $1 \times 10^{-4}$ joule +0 joule $-0.5 \times 10^{-4}$ joule.

With each successive strike of the hammer, the attenuated field would decrease until a constant value was reached. Approximately 100 strikes were required to achieve the minimum. The value would not change with additional hammer strikes. The change of attenuated field as a function of the number of hammer strikes is shown for the Cryoperm10 sample in Figure 1.

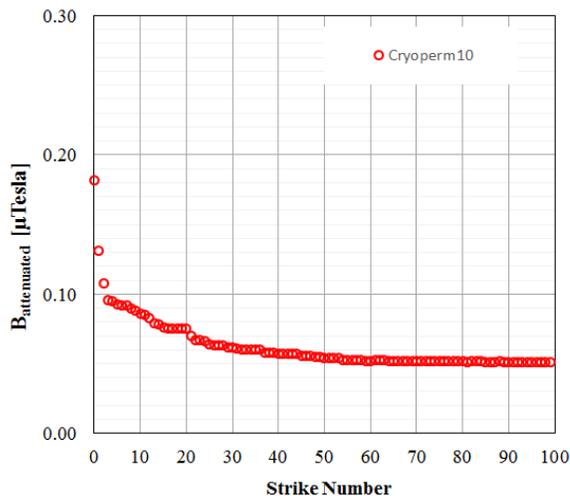

Fig 1. Hammer strike sequence for the Cryoperm10 sample.

Average values for measured attenuated field and calculated relative permeability (μ$_r$) for the cases of before and after mechanical perturbation are listed in Table V. The applied axial field is 50.0 μTesla.

TABLE V
EFFECT OF STIMULATION ON ATTENUATION

| SAMPLE | B$_{AXIAL\ before}$ (μTESLA) | B$_{AXIAL\ after}$ (μTESLA) | μ$_{r\ before}$ | μ$_{r\ after}$ |
|---|---|---|---|---|
| 1115 | 0.081 | 0.069 | 98,500 | 116,000 |
| 1184 | 0.168 | 0.110 | 47,500 | 75,500 |
| A4K | 0.090 | 0.062 | 88,500 | 128,000 |
| Cryoperm10 | 0.178 | 0.101 | 44,000 | 77,500 |

The change to attenuated field due to mechanical stimulation was found to be a reversible effect. Sequential measurements made on all samples are shown in Figures 2 through 5. The sequence for each sample is: 1. Measurement with remanent field in sample adjusted to zero, 2. Stimulate, 3. Reverse axial orientation of sample, 4. Stimulate, 5. Reverse, 6. Stimulate, 7. Reverse, 8. Stimulate, 9. Reverse, 10. Stimulate, 11. Reverse. The even numbers are measurements where the sample has been stimulated within the applied field in which the measurement was made. For odd numbers greater than one, the axial orientation of the sample within the constant applied field has been reversed from the time of the mechanical stimulation.

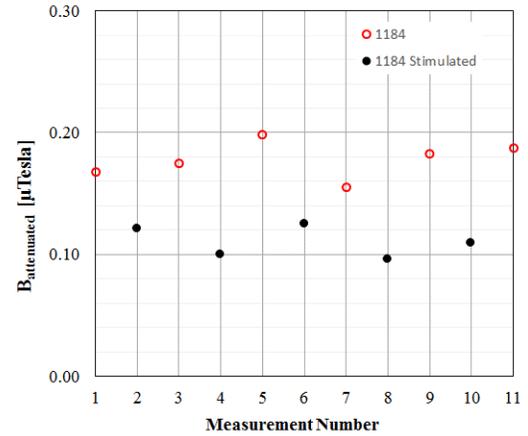

Fig 3. Sequence of measurements on the 1184 sample.

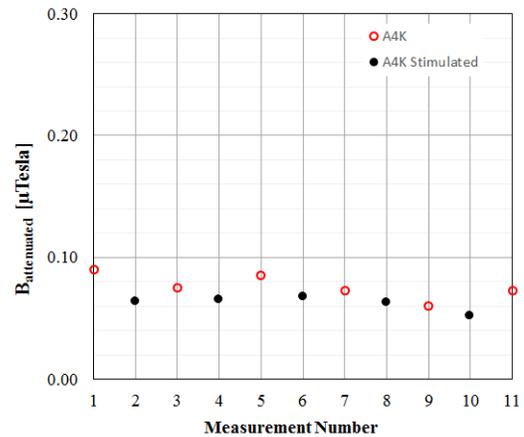

Fig 4. Sequence of measurements on the A4K sample.

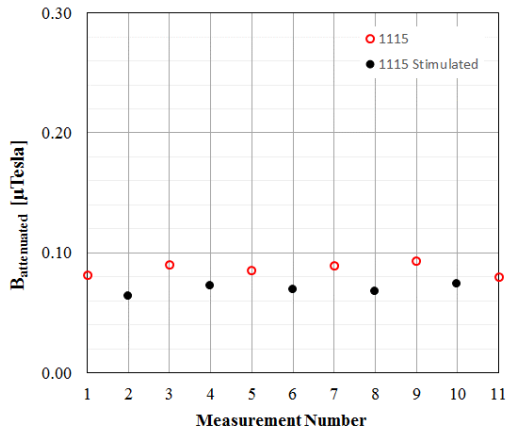

Fig 2. Sequence of measurements on the 1115 sample.

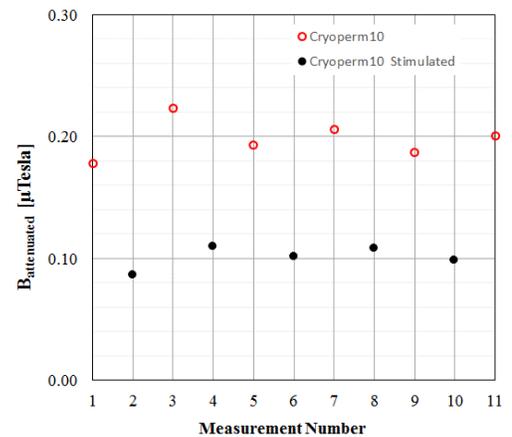

Fig 5. Sequence of measurements on the Cryoperm10 sample.

When hammer strikes with a larger amount of kinetic energy were applied, the attenuated field would decrease to a new, smaller equilibrium value with higher effective permeability. It is anticipated that with increasing energy level the effects will cease to be reversible at some point. Within the reversible limits, we have, in effect, a mechanically stimulated hysteresis function for a given applied magnetic field.

## VI. Discussion

When magnetic shielding material is removed from the annealing furnace it possesses two different forms of internal structure that are dependent on the thermal treatment conditions: crystalline grains and magnetic domains. It is likely that the samples in this study have not been subjected to forces sufficient to cause re-ordering of grain boundaries or to introduce significant permanent internal stress. We are observing magnetic domain re-arrangement resulting from mechanical stimulation and applied ambient magnetic field.

The small amount of mechanical stimulation given to the samples is enough to allow the magnetic domain walls to move in a way that allows lessening of the total Gibbs free energy of the system in the applied magnetic field [4]. This translates to a larger integrated magnetization within the sample in alignment with the applied field. This type of metastable equilibrium is common in nature. A ball on a perfectly level table will remain stationary under normal conditions. If someone hits the table with a hammer, there is a chance that the ball will roll off the table. If someone hits the table with a hammer one hundred times, there is a larger probability that the ball will roll off the table. When the ball falls from the table, the free energy of the system is lowered.

The amount of energy imparted to the samples in this study is small, an amount that can be transmitted to magnetic shields even after they are installed in cryomodules. If cryomodule shields behave like the samples, then vibrational energy from handling and transportation can cause the effect seen in this test. Magnetic domain walls would be arranged according to the magnetic and mechanical history of the cryomodule's journey to its final location. The result is not likely to be the optimum arrangement for permeability of the shield in-situ.

Do cryomodule shields behave like the samples? By the time that cryomodule shields are installed they have usually been subjected to enough force to modify the grain structure and to cause permanent internal stress. Previous measurements indicate that the room temperature relative permeability of Cryoperm10 shields for Fermilab International Linear Collider (ILC) cryomodules is not higher than 24,000 [5]. Measurements on Cryoperm10 shields for the Linac Coherent Light Source-II (LCLS-II) prototype cryomodule indicate a relative permeability of 10,000. These values are significantly lower than for any sample of this study.

Could in-situ mechanical stimulation be used to increase the relative permeability of cryomodule shields? So far, the answer seems to be, no. Sections of ILC and LCLS-II magnetic shields have been tested in a similar manner to the samples in this study. No change was observed in effective permeability with applied mechanical stimulation, even with energy reaching 1.5 joules per Kilogram. Speculation is that this is due to grain damage and stress concentration in the ILC shields. The profound effect of stress on permeability of cryogenic magnetic shielding materials is demonstrated in reference [6].

Where does this leave the SRF magnetic shield designer? How does one answer the question "What is the permeability?" when designing shields? The answer is that the permeability is unknown unless attenuation is measured in-situ under the exact circumstances that the shield will be used. The conservative designer will use a reasonable approximation of the worst case permeability for the material that he is using, and include suitable cryogenic magnetometry in the cryomodule to measure the field at the cavities for verification. An assumed relative permeability of 10,000 appears to be a reasonable choice to use for modeling until suitable techniques can be developed to preserve the furnace treated permeability of shielding material.

## VII. Conclusion

Specimens of magnetic shielding materials prepared for cryogenic use show large changes in their relative permeability when subjected to small mechanical shock. The effect is reversible, causing no permanent change to the magnetic properties. Limited mechanical shock increases the effective permeability and therefore the ability of the material to attenuate applied field.

The results of this study apply to carefully prepared and handled samples only. They do not represent practically realizable permeability for real world magnetic shields. Real world performance is likely to be significantly worse than the attenuation factors measured here.

## VIII. Future work

The samples of this study will be tested at 77K and 4.2K. Work will continue toward better understanding of the best realizable permeability for cryomodule magnetic shields, with special attention applied to identifying where the loss of permeability occurs from the annealing furnace to the cryomodule.

Acknowledgment

The author thanks Derrick Vance of Ad-Vance magnetics for providing very interesting sample specimens and for their preparation. Thanks to Sam Posen for thoughtful and helpful comments on this report.

APPENDIX

Magnetic viscosity effects were observed in all the samples of this study. This was especially evident when the sample cylinders were reversed in the 50 µTesla field. The time constant for decay of the attenuated field was approximately 10 seconds. All measured values in this report are taken after the attenuated field has approached its asymptotic limit (6 time constants). The materials of this study do not exhibit magnetic superviscosity effects as does the Metglass of reference [4].